\documentstyle[12pt]{article}
\evensidemargin \oddsidemargin
\newcommand{\bm}{\bibitem}
\def \z {(\sqrt{-p^2} /\mu )}
\def \y {(\sqrt{-p^2} / T)}
\begin{document}
\begin{center}
{\Large{\bf Absence of a second order phase transition in 
 $\lambda\phi ^4$ theory}}\\

\vspace{1cm}

S. Mallik\footnote{E-mail: mallik@tnp.saha.ernet.in}
 and Krishnendu Mukherjee\footnote{E-mail: krish@tnp.saha.ernet.in}\\

\vspace{.5cm}

Saha Institute of Nuclear Physics, 1/AF, Bidhannagar, Calcutta - 700064,
India\\

\end{center}

\vspace{1.5cm}

\noindent {\bf Abstract} 

We calculate the self-energy at finite temperature in scalar $\lambda\phi ^4$
theory to second order in a modified perturbation expansion. Using the 
renormalisation group equation to tame the logarithms in momentum, it gives
an equation to determine the critical temperature.
Due to the infrared freedom of the theory, this equation is satisfied,
irrespective of the value of the temperature. We conclude that there is no
second order phase transition in this theory.

\vspace{1.5cm}
\newpage
\section{Introduction} 
In this work we examine the possibility of a second order phase transition in
the scalar field theory with the Lagrangian density
\begin{equation}
{\cal L} = {1\over 2}(\partial_{\mu}{\phi} )^2 - {1\over 2} m^2\phi ^2 -
           {\lambda\over 4!}\phi ^4\label{lag1},
\end{equation}
where $m^2 < 0$. We have to calculate the effective
mass at finite temperature in the symmetric phase. As the 
temperature is lowered, this mass may vanish
at a definite temperature, indicating 
a second order phase transition [1-3].

A general problem here is the breakdown of usual perturbation expansion when
powers of temperature compensate for powers of coupling constant. In the
scalar theory it is due to the generation of the thermal mass, which is generally
taken into account by a summation over the so-called ring or daisy diagrams
[2,4].
An alternative and more consistent method is to modify the perturbation 
expansion by adding a (temperature dependent) mass term in the free Lagrangian
and a compensating counterterm in the interaction [3,5-7]. Thus we rewrite
(1) as
\begin{equation}
{\cal L} = {1\over 2}(\partial_{\mu}\phi )^2 - {1\over 2} M^2 \phi ^2
          -{\lambda\over 4!}\phi^4 - {1\over 2}(m^2 - M^2)\phi^2\label{lag2}
\end{equation}
The perturbation series will be a joint expansion in powers of $\lambda$ and
$(m^2 - M^2)$. Although $(m^2 - M^2)$ may not be small, it will appear
in a combination which would be small. The parameter M becomes the
effective mass, if we require perturbative corrections to the mass term
(at zero momentum) to be zero. In general all these methods of taking the 
thermal mass into account are also equally effective in removing the  
infrared divergences of a massless field theory at finite temperature.

To be specific, let us denote the self-energy at momentum $p_{\mu}$ and temperature
$T$, calculated upto a certain order of our perturbation expansion, 
by $\Sigma (p^2,M,T)$. ( We suppress its 
dependence on $m$, $\lambda$ and the renormalisation scale $\mu$ to be introduced 
below.) Then the effective mass $M$ is obtained by solving 
\begin{equation}
\Sigma (p^2=0,M,T) = 0.
\end{equation}
Formally the critical temperature $T_c$ of a second order phase transition is attained
when $M$ goes to zero,
\begin{equation}
\Sigma (p^2=0,M=0,T_c) = 0.
\end{equation}

The other problem, specific to our analysis, is the new infrared divergence
encountered in (4): If we first set $p_{\mu}=0$ and calculate $\Sigma$
perturbatively beyond the first order, 
there arises powers of $ln M^2$ (multiplied with $T^2$ or $m^2$) [5].
Thus $\Sigma (p^2=0,M,T)$ diverges logarithmically as $M\rightarrow
{0}$ and (4) is not meaningful.
In the same way, if we first set $M=0$, then $\Sigma (p^2, M=0, T)$ contain 
powers of $\ln p^2$ and again we cannot reach the limit in (4).

As the pure scalar theory is infrared free, it is possible to sum these 
infrared divergences in perturbation expansion
by the renormalisation group equation. The equation we use is identical
to the one at zero temperature.\footnote{We are not using the scheme where
the parameters in the Lagrangian are renormalised not only at a momentum
scale $\mu$ but also at a temperature $T_{0}$, say, as in Ref.[8].}
But due to the presence of an additional scale,
viz, the temperature, the solution generates new logarithmic singularities. 
They are, however, tamed by multiplication with 
the running coupling constant. In sec.4 we
shall set $M =0$ and sum the leading logarithms in $p^2$.

\section{Self energy to second order}

 To make the cancellation of ultraviolet divergences explicit in the calculation,
we augment the Lagrangian density (1) with the renormalisation counterterms
\begin{equation}
{\cal L}_{ct} = {1\over 2} A (\partial_{\mu}\phi)^2 - {1\over 2} m^2 B \phi^2
              - {\lambda\over 4!} \mu^{2\epsilon} C \phi^4\label{lag3}
\end{equation}
where A, B, and C are the divergent coefficients. In the minimal subtraction 
scheme of dimensional regularisation in dimension $d$, 
 they are, to order $\lambda^2$ [9],
\begin{equation}
A = - {{\hat{\lambda}}^2\over 24\epsilon}, \qquad 
B = {\hat{\lambda}\over 2\epsilon} + {\hat{\lambda}}^2 ({1\over 2\epsilon^2} 
- {1\over 4\epsilon}), \qquad
C = {3\hat{\lambda}\over 2\epsilon} 
+ {3\over 4}\hat{\lambda} ({3\over {\epsilon}^2} 
- {2\over \epsilon}), 
\end{equation} 
where $\hat{\lambda} = \lambda /{ 16\pi^2}$, $\epsilon = 2-d/2$ and
$\mu$ is the renormalisation scale. Corresponding to the modified scheme          
(2), the total Lagrangian splits as 
\begin{equation}
{\cal L} + {\cal L}_{ct} = {1\over 2}(\partial_{\mu}\phi)^2 - {1\over 2} M^2\phi^2
 - {\lambda\over 4!} \mu^{2\epsilon}\phi^4 + {1\over 2} A (\partial_{\mu}\phi)^2 
-{1\over 2} M^2 B\phi^2 - {\lambda\over 4!}\mu^{2\epsilon} C\phi^4  \\
                      -{1\over 2}(m^2 - M^2) (1+B)\phi^2 \label{lag4}
\end{equation}
Because of our use of the minimal subtraction scheme, where the renormalisation
constants are mass independent, the Lagrangian (7) is, except for
the last term, the same as the sum of Lagrangians (1) and 
(5) with $m^2$ replaced by $M^2$. The last term incorporates its
own renormalisation counterterm.

We use the real-time formulation of the finite temperature field theory 
[10], where the thermal propagator becomes a $2\times 2$ matrix. 
Working below threshold, we need only calculate the $11$-component of the
$\Sigma$-matrix, which we have already denoted by $\Sigma$.

The self-energy diagrams to second order are shown in Fig.1. Digrams (a) 
and (c) are specific to our modified perturbation expansion. It will be 
observed that, except for the diagram (e), all others are independent of
momentum and so particularly simple to evaluate [2-7,9,11]. To begin with, 
we do not restrict either the external momentum $p_{\mu}$ or the effective mass
$M$ to zero. Denoting these contributions by $-i\Sigma$ , the diagrams (a)
and (b) contribute respectively
\begin{equation}
\Sigma_{(a)} = (m^2 - M^2)(1+B), \qquad  \Sigma_{(b)} = G(M^2), 
\end{equation}
where
\begin{equation}
G(M^2) = {\lambda\over 2}\left (i\int {(dl)\over {k^2 - M^2 + i\epsilon}}
 + 2\pi \int
(dl) n(\omega_l) \delta (l^2 - M^2) \right ) \label{B}
\end{equation}
$(dl)$ standing for $d^d l/{ (2\pi)^d}$ . Isolating the divergent piece, we
write
\begin{equation}
G(M^2) = -{{\hat{\lambda}} M^2\over 2\epsilon} + \bar{G}(M^2).
\end{equation} 
$\bar{G}$ has the high temperature expansion (${M/T}$ small) [2],
\begin{equation}
\bar{G}(M^2) = {\lambda\over 2}\{ {T^2\over 12} - {MT\over 4\pi} + 
{M^2\over 8\pi^2}(\ln{T\over {\mu}} + const.)  + O({M^4\over T^2})\} \label{C}
\end{equation}
The two integrals in (9) separately contain $M^2ln M^2$ terms, which
cancel out in the sum $\bar{G}$. 
Using the mass derivative formula [12], diagrams (c) and (d)
can be evaluated as
\begin{equation}
\Sigma_{(c)} = (m^2 - M^2) (1+B) {\partial\over \partial{M^2}}G, \qquad
\Sigma_{(d)} = G {\partial\over \partial{M^2}}G
\end{equation}
 
We next evaluate the diagram (e) of Fig.1,
\begin{eqnarray}
\Sigma_{(e)}&=& -{\lambda^2\over 6}\int\int (dk)(dl) {1\over {k^2 - M^2 + 
i\epsilon}} {1\over {l^2 - M^2 + i\epsilon}} {1\over {(p - k - l)^2 - M^2 + 
i\epsilon}}\nonumber \\
 &+& i{\lambda^2\over 2}\int (dk) 2\pi \delta{(k^2 - M^2)} 
n(\omega_k) \int (dl) {1\over {l^2 - M^2 + i\epsilon}}
{1\over {(p - k - l)^2 - M^2 + i\epsilon}} \nonumber\\ 
 &+& {\lambda^2\over 2} \int \int (dk) (dl) (2\pi)^2
 \delta{(k^2 - M^2)}\delta{(l^2 - M^2)} n(\omega_k)
 n(\omega_l) {1\over {(p - k - l)^2 - M^2 + i\epsilon}}.    
\end{eqnarray}
The first term corresponds to the zero temperature contribution. Separating
the divergent pieces, it becomes $(\hat {M}^2=M^2/{4\pi {\mu^2}})$,
\begin{equation}
{\hat{\lambda}}^2\{ {M^2\over 4\epsilon^2} + {M^2\over 2\epsilon}
({3\over 2} - \gamma - \ln{\hat{M}}^2) - {p^2\over 24\epsilon}\}
+F_1{(p , M , \mu)}.    
\end{equation}
Similarly the second term may be written as 
\begin{equation}
-{{\hat{\lambda}}\over \epsilon}\{ \bar{G} - {{\hat{\lambda}}M^2\over 2}
(\ln{\hat{M}}^2 + \gamma - 1)\} + F_2{(p , M , \mu , T)}
\end{equation}
The third term is free from divergence, to be denoted by $F_3{(p, M, T)}$.
The finite pieces $F_1$, $F_2$ and $F_3$ are complicated functions of $p^2$
and $M^2$ and will be evaluated in sec.3 below in the required limit.

Finally the diagrams (f) are due to the renormalisation counterterms,
\begin{equation}
\Sigma_{(f)} = M^2 \{ {{\hat{\lambda}}\over 2\epsilon} + {\hat{\lambda}}^2 
({1\over 2\epsilon^2} - {1\over 4\epsilon}) \} + 
{{\hat{\lambda}}^2\over 24\epsilon}p^2
              + {{\hat{\lambda}}^2\over 2\epsilon}
 M^2{\partial\over \partial{M^2}}G + {3\hat{\lambda}\over 2\epsilon}G.
\end{equation}
One may now check that all the divergent pieces, belonging to the different
diagrams, 
cancel out. The complete self-energy to second order in our modified 
perturbation expansion becomes
\begin{equation}
\Sigma(p^2, M, T) = (m^2 - M^2) + {\bar{G}}
  + (m^2 - M^2 + \bar{G})
{\partial\over \partial{M^2}}{\bar{G}} + F\label{D}
\end{equation}
where
\begin{equation}
F(M) = F_1{(p, M, \mu)} + F_2{(p, M, \mu, T)} + F_3{(p, M, T)}
\end{equation}
Noting the second term in (11) for $\bar{G}$, we see that the third term in
(17) for $\Sigma$ has a linear divergence as $M\rightarrow 0$.  

\section{Infrared divergence}

As already mentioned in the Introduction, the familiar infrared divergence of
a massless boson theory
at finite temperature is cured by incorporating the thermal mass 
into the mass term of the free propagator. While such an effective mass serves as an 
infrared cut off in general, it itself tends to zero as one approaches the critical
temperature, giving rise to the infrared problem in (17). Being 
analogous to that of an originally massless theory, this infrared
singularity may be removed by summing over the  
daisy diagrams (Fig.2)
of our modified perturbation expansion.
\footnote{It should be noted that there is no double counting
here. If we had taken the thermal mass into account by summing over the
daisy diagrams instead of modifying the perturbation expansion, the daisy
diagrams of Fig.2 would correspond to 
a kind of superdaisy diagrams in that scheme.}
Again using the mass derivative formula, this sum is easily seen to be 
a Taylor series [4],
\begin{equation}
\Sigma_{(daisy)} = \sum_{n=0}^{\infty} {1\over n!}(m^2 - M^2 +
\bar{G}(M^2))^n ({\partial\over \partial{M^2}})^n {\bar{G}(M^2)}
                 = \bar{G} (m^2 + \bar{G}(M^2))
\end{equation}
which is to replace the second and third terms in (17).

Having gotten rid of the divergence, we may set $M=0$ in (17)
to get
\begin{equation}
\Sigma(p^2, M=0,  T) = {\cal M}^2 - {{\lambda T}\over 8\pi}{\cal M} 
                      + \hat{\lambda} {\cal M}^2
 (\ln{T\over \mu} +
const) + F(M=0),
 \end{equation}
where ${\cal M}^2= m^2 +\lambda T^2/{24}$. The logarithmic term above
multiplying $m^2$ and $T^2$ may be
checked to be identical to those in (17). The $F_i$'s are of
course, finite at $M=0$, as long as $p^2 \neq 0$.
Their logarithmic pieces can now be obtained
without difficulty. The zero temperature contribution gives
\begin{equation}
F_1 (p^2, M=0, \mu) = {{\hat{\lambda}}^2\over 6} p^2 (\ln{\z} + const)
\end{equation}
It remains to evaluate the other two finite pieces given by
\begin{equation}
F_2{(p^2, M=0, T, \mu)} = \lambda{\hat{\lambda}}
\pi\int (dk) \delta{(k^2)} n(\omega _k) \ln{\left (-{{(p - k)}^2\over
{\mu}^2}\right )},
\end{equation}
\begin{equation}
 F_3{(p^2, M=0, T)} = {\lambda ^2\over 2}\int
(dk)\int (dl) (2\pi)^2 n(\omega_k) n(\omega_l)\delta (k^2)\delta (l^2)
{1\over {(p - k - l)^2 + i\epsilon}}
\end{equation}
To evaluate these for space-like $p^2$,
we may simplify the calculation by taking $p=(p_0 , \vec{0})$, $p_0$
imaginary , and re-expressing the result finally in terms of $-p^2 >0$. In
this way we get, 
\begin{eqnarray} 
F_2 &=& {\lambda{\hat{\lambda}}\over 24}
T^2\{ \ln{\z} + \ln{T/ \mu} + const \}\\ 
F_3 &=& {\lambda{\hat{\lambda}}\over 16} T^2\{ \ln{\y} + const\} 
\end{eqnarray}
Having calculated $F(M=0)$, we finally get, 
\begin{eqnarray} 
\Sigma ( p^2, M=0,T) &=& {\cal M}^2  - \lambda T {\cal M}/{8\pi}
  + {{\hat{\lambda}}\lambda\over 8}T^2 (\ln{\z} -{1\over 6} \ln{\y})\nonumber \\  
                     &+& {\hat{\lambda}}m^2 (\ln{\z} - \ln{\y})
 + {{\hat{\lambda}}^2\over 6}p^2 \ln{\z} 
+ O(\lambda ^2 T^2). 
\end{eqnarray}
\section{Renormalisation group method}
Because of the presence of the $ln{\z}$ in the expression (26) for 
$\Sigma$, we cannot approach zero momentum. We have to improve the one-particle 
irreducible, two point function,
\begin{equation}
\Gamma^{(2)}(p, \lambda , m, \mu , T) = -i( p^2 - \Sigma (p^2,M=0,T)),
\end{equation}
by summing the leading logarithms of the perturbation series. This may be conveniently
done by using the solution of the renormalisation group equation,     
\begin{equation}
\left (\mu {\partial\over \partial\mu} + \beta (\lambda ){\partial\over \partial\lambda} + 
m\gamma_m (\lambda){\partial\over \partial{m}} - \gamma_{\phi} (\lambda)
\right ) 
\Gamma^{(2)} (p, \lambda, m, \mu, T) = 0 \label{F}
\end{equation}
The coefficients in this equation are calculated from the renormalization
counterterms. To lowest order, they are
\begin{equation}
\beta = 3\lambda\hat{\lambda},~~~\gamma_m = {{\hat{\lambda}}\over 2},~~~
\gamma_{\phi} = {1\over 6}{\hat{\lambda}}^2.
\end{equation}

Let us first check that the $\mu$ dependence of $\Gamma^{(2)}$ as calculated above
indeed satisfies the renormalization group equation [13]. We write
$\Gamma^{(2)}$ as series in powers of $\lambda$, $(z = \z)$,
\begin{eqnarray}
i\Gamma^{(2)} &=& T^2 \sum_{n=1} a_n(z)\lambda^n + m^2(-1+\sum_{n=1}
b_n(z)\lambda^n) + p^2(1+\sum_{n=2}d_n(z)\lambda^n)\nonumber\\
              & & + \rm{\; terms \; independent \; of \; z}.
\end{eqnarray}
Substituting this in (28) and equating like powers of $\lambda$ for
each of these series to zero, one gets simple differential equations
in $\mu$ or $z$ with the solutions,
\begin{eqnarray} 
a_1(z)&=&c_1,~~~a_2(z)={3c_1\over { 16\pi^2}}\ln z + c_2,\nonumber\\
b_1(z)&=&-{1\over {16\pi^2}} \ln z + c_3,~~d_2(z)=-{1\over {6({16\pi^2})^2}}\ln z + c_4,
\end{eqnarray}
where $c_i$'s are constants. From (26) we see that they are indeed satisfied.

The renormalization group equation used here is the same
as the one of conventional (zero tempemperature) field theory. Being determined 
by the short distance properties of the theory, the coeffiecients $\beta$,
$\gamma_{m}$ and $\gamma_{\phi}$ do not know of any temperature, at least
in the minimal subtraction scheme. It is only through the set of constants
$c_i$ in (31) that temperature enters $\Gamma^{(2)}$.

The solution of the renormalization group equation relates $\Gamma^{(2)}$
at two different scales. Let us consider $\Gamma^{(2)} (sP, \lambda, m, \mu,
T )$ and take $s\rightarrow 0$ at the end $(P\neq 0)$. Then 
we write the solution as 
\begin{equation}
\Gamma^{(2)} (sP, \lambda, m, \mu, T) = s^2 \Gamma^{(2)}(P, \lambda, 
{m\over s}, {\mu \over s}, {T\over s})   = s^2 Z(s) \Gamma^{(2)} (P,\bar{\lambda}
, {{\bar{m}}\over s}, \mu, {T\over s}),
\end{equation}
where we use dimensional analysis in the intermediate step. While $\lambda$ and
$m$ are renormalised at the scale $\mu$, $\bar{\lambda}$ and $ \bar m$ are 
referred to the scale $\bar {\mu} = s\mu$, with the relations
\begin{equation}
\bar {\lambda} = {\lambda \over {1-3\hat {\lambda} \ln s }},\qquad
\bar m= m \left ( {\bar{\lambda}\over \lambda} \right )^{1\over 6}.
\end{equation}
Z(s) given by
\begin{equation}
Z(s) = exp\big [ -\int_{\lambda}^{\bar{\lambda}}{\gamma
(\lambda^{\prime})\over \beta(\lambda^{\prime})} d\lambda^{\prime}\big ] 
 = exp[-{(\bar\lambda - \lambda)/ {288\pi^2}}],
\end{equation}
tends to a constant as $s\rightarrow 0$.
Inserting our evaluation (26,27) of $\Gamma^{(2)}$ in the right hand side of
(32) and choosing $P_{\mu}$ and $\mu$ such that $\sqrt{-P^2} =\mu =T$ to get 
rid of the (finite) logarithms, we get
\begin{eqnarray}
i Z(s)^{-1}\Gamma^{(2)}(sP, \lambda, m, \mu, T) &=& s^2 P^2 - {\bar m}^2 - 
{\bar{\lambda} T^2\over {24}} +{\bar{\lambda} T\over {8\pi}}\left (
{\bar{m}}^2 +{{\bar{\lambda}} T^2\over {24}}\right )^{1\over 2}\nonumber \\
                                     &+& {\bar{\lambda}\ln s\over {16
\pi^2}}\left ({\bar {m}}^2 +{\bar{\lambda}T^2\over {48}}\right )+O(\lambda ^2
T^2) .   
\end{eqnarray}
We see that although terms with $ln s$ arising from $ln \z$ are absorbed
in the
parameters $\bar\lambda$ and $\bar m$, new $ln s$ terms arise from
$ln \y$ present in $\Sigma$. Though they cannot lead to a
logarithmic divergence as $s\rightarrow 0$, because of multiplication
with $\bar\lambda$ (and $\bar m$), we see that there will arise terms
proportional to $\bar\lambda$ and $\bar m$ from each order of the
perturbation expansion. They will form a series in powers of
$\bar\lambda \ln s/{ 16\pi^2}\rightarrow -{1/3}$ as $s\rightarrow 0$.

The point to observe is that each term in the expression (35) for
$\Gamma^{(2)}$goes to zero as the momentum tends to zero.
Retaining only the first two leading terms for small $s$, we have
\begin{eqnarray}
\Sigma(p^2=s^2 P^2, M=0,  T) &=& -i\Gamma^{(2)}(sP, \lambda, m, \mu,
T)\nonumber \\
                             &=&Z[ {\bar m}^2(1+{1\over3}+\cdots) 
+ {\bar\lambda
T^2\over24}(1+{1\over6}+ \cdots) + O(ln s)^{-{7\over 6}}],
\end{eqnarray}
where dots denote contributions beyond the second order.
\section{Conclusion}
We see that equation (4), which is supposed to give a definite value for the
critical temperature, is trivially satisfied,
independently of the value of the temperature. We conclude that there
is no proper second order phase transition in the pure scalar field theory.
It is not difficult to see that the same conclusion would have been reached
if we had
first put $p_{\mu}=0$ and then summed the resulting series in $\ln M^2$ by the
renormalisation group equation. Clearly our result is a direct consequence
of the infrared freedom of the scalar field theory in dimension $d=4$.

 In dimensions $d<4 (\epsilon >0)$, of interest in condensed matter physics,
the scalar field theory develops a stable, non-zero infrared fixed point, at
least in the perturbative calculation [14]. There one gets a finite critical
temperature for a second order phase transition. But in condensed
matter physics, the scalar field theory is used phenomenologically and
predicts behaviour of correlation functions at or near the critical point
but not the value of the critical temperature itself.

Going beyond the scalar field theory, we may say that there cannot also be a
second order phase transition in any other system whose contents are described by
an infrared free theory, e.g. the combined system of the Higgs and QED.

We wish to thank Professors P. Aurenche,
S. Gupta, R. Jackiw and M. Shaposhnikov for
discussions on related topics.

\end{document}